\documentclass[prd,preprint,eqsecnum,nofootinbib,amsmath,amssymb,preprintnumbers,tightenlines,longbibliography]{revtex4-1}

\usepackage{afterpage}
\usepackage{mathrsfs}
\usepackage{graphicx}
\usepackage{bm}
\usepackage{paralist}
\usepackage[colorlinks,linkcolor=blue,citecolor=blue]{hyperref}

\def\Tf{\tau_{\rm final}}
\def\e{{\epsilon} }

\def\c{{\mathcal C}}

\def\psie{\Phi}
\def\phis{\phi }

\def\ls2{{\ell_s^2}}
\def\x{{\bm x}}
\def\dd{{\rm d}}

\newcommand\bga{\begin{align}}
\newcommand\nda{\end{align}}

\def\x{{\bm x}}

\def\p{{\bm p}}

\def\half{{\textstyle\frac{1}{2}}}

\def\st{\begin{equation}}
\def\stp{\end{equation}}
\def\bg{\begin{eqnarray}}
\def\nd{\end{eqnarray}}
\def\Eq#1{eq.~(\ref{#1})}

\def\Fig#1{Fig.~\ref{#1}}
\def\Sect#1{Section~\ref{#1}}
\def\Ref#1{Ref.~\cite{#1}}

\def\llangle{\left\langle}
\def\rrangle{\right\rangle}
\def\sllangle{\llangle }
\def\srrangle{\rrangle }

\def\Bigdlangle{\Big\langle\!\!\Big\langle}
\def\Bigdrangle{\Big\rangle\!\!\Big\rangle}
\def\Dlangle{\left\langle\!\left\langle}
\def\Drangle{\right\rangle\!\right\rangle}

\def\nott#1{\setbox0=\hbox{$#1$}                
   \dimen0=\wd0                                 
   \setbox1=\hbox{/} \dimen1=\wd1               
   \ifdim\dimen0>\dimen1                        
      \rlap{\hbox to \dimen0{\hfil/\hfil}}      
      #1                                        
   \else                                        
      \rlap{\hbox to \dimen1{\hfil$#1$\hfil}}   
      /                                         
   \fi}                                         %

\advance\parskip 1.9pt
\advance\voffset -0.2in

\def\st{\begin{equation}}
\def\stp{\end{equation}}
\def\bg{\begin{eqnarray}}
\def\nd{\end{eqnarray}}

\begin{document}

\title{Non linearities in the harmonic spectrum of heavy ion collisions
with ideal and viscous hydrodynamics}

\author{D.~Teaney}
\email{derek.teaney@stonybrook.edu}
\author{L.~Yan}
\email{li.yan@stonybrook.edu}
\affiliation
    {%
    Department of Physics \& Astronomy,
    Stony Brook University,
    Stony Brook, NY 11794, USA
    }%

\begin{abstract}
   We determine the non-linear hydrodynamic response to geometrical fluctuations in heavy ion collisions using ideal and viscous hydrodynamics.  This response is characterized with a set of non-linear response coefficients that determine, for example, the $v_5$ that is produced by an $\epsilon_2$ and an $\epsilon_3$.  We analyze how viscosity damps both the linear and non-linear response coefficients, and provide an analytical estimate that qualitatively explains most of the trends observed in more complete simulations.  Subsequently, we use these non-linear response coefficients to determine the linear and non-linear contributions  to $v_1$, $v_4$ and $v_5$.  For viscous hydrodynamics the non-linear contribution is dominant for $v_4$, $v_5$ and higher harmonics.  For $v_1$, the non-linear response constitutes an important $\sim 25\%$ correction in mid-central collisions. The non-linear response is also analyzed as a function of transverse momentum for $v_1$, $v_4$ and $v_5$.  Finally, recent measurements of correlations between event-planes of different harmonic orders are discussed in the context of non-linear response.

\end{abstract}

\date{\today}

\maketitle

\section{Introduction} 

The goal of the RHIC and the LHC heavy ion programs is to produce and to characterize
the Quark Gluon Plasma (QGP), a prototype for non-abelian plasmas.  One of the
best ways to  understand the transport properties of the experimentally
produced plasma is through anisotropic flow
\cite{Voloshin:2008dg,Teaney:2009qa,Kolb:2003dz}.  
In a heavy ion collision the nuclei pass through each
other, and the resulting energy density in the transverse plane  
fluctuates in coordinate space from event to event.  
If the mean free path is short compared to the system size,
the produced plasma will respond as a fluid to the pressure gradients and  convert
these coordinate space  fluctuations  to long range momentum space correlations
between the produced particles.  In the last two years it was gradually
realized \cite{Takahashi:2009na,Sorensen:2010zq,Ma:2006fm} that all of the long range momentum-space correlations known 
 colloquially as the ``ridge" and ``the Mach cone" are manifestations
of this collective flow \cite{Alver:2010gr,Luzum:2010sp}. 
This realization gave rise to a large variety of flow observables
which provide an unprecedented experimental check of the overall
correctness of the  hydrodynamic  picture of heavy ion events \cite{Alver:2010gr,Teaney:2010vd,Bhalerao:2011yg,ATLASCorrelations}. 
Further, different observables have different sensitivity to the shear
viscosity  of the plasma \cite{Alver:2010dn}, and therefore a global analysis of flow can provide cross-correlated constraints on  $\eta/s$.

One of the most direct measurements is the harmonic spectrum of the produced
particles.  The final state momentum spectrum for each event  can be expanded
in harmonics
\st
\label{spectra_intro}
 \frac{dN}{ d\phi_\p  } 
= \frac{N}{2\pi}\,
\left(1 + 2\sum_{n=1}^{\infty} 
  v_n \cos(n \phi_\p - n\Psi_n)  
 \right) \, , 
\stp
where $\phi_\p$ is the azimuthal angle of the produced particles 
and $\Psi_n$ is the event plane angle\footnote{
Following tradition, we have expanded the particle distribution 
in terms of cosines and phases $\Psi_n$ rather than 
cosines and sines.}. 
 The averaged square of
these harmonics, {\it i.e.} $\Dlangle v_n^2 \Drangle$,  can be measured  experimentally by studying
two particle correlations \cite{Voloshin:2008dg}.
There is strong experimental and theoretical evidence  that the harmonic
coefficients, $v_2$ and $v_3$, are to a good approximation linearly proportional to the deformations in
the initial energy density in the transverse plane.
For example,  the experimental ratio $\Dlangle v_3^2 \Drangle/\Dlangle v_2^2 \Drangle$
closely follows the geometric  deformations $\Dlangle \epsilon_3^2 \Drangle/
\Dlangle \epsilon_2^2 \Drangle $ as a function of 
centrality \cite{Alver:2010gr}.  Event-by-event simulations with ideal hydrodynamics reproduce this trend,
and show that the event plane 
angles $\Psi_2$ and $\Psi_3$
are strongly correlated with the angles of the initial deformations \cite{Qiu:2011iv}.

However, in an insightful paper Gardim {\it et al} \cite{Gardim:2011xv}
studied the correlation between higher harmonics, $v_4$ and $v_5$,  and the
initial spatial deformations within ideal hydrodynamics. This work explained
and quantified the extent to which the higher harmonics such as $v_4$ and $v_5$
arise predominantly from the non-linearities of the medium response.  For
example, for mid-central collisions the observed $v_5$ is predominantly a
result of the interactions between $v_2$ and $v_3$.  This work was motivated in
part by previous event-by-event simulations by Heinz and Qiu \cite{Qiu:2011iv}
which showed  that  $\Psi_4$ and  $\Psi_5$ are uncorrelated with the fourth and
fifth harmonics of the   spatial deformation. Based on the centrality
dependence  of this decorrelation, these authors anticipated (but did not
quantify) the importance of $v_2$-$v_3$ mode-mixing in determining $v_5$. 

The goal of this work is to systematically characterize the non-linear response
of the medium.  First, in \Sect{nonlin} we introduce a set of non-linear response
coefficients, and describe how these coefficients can be used in conjunction
with a Glauber model to determine $\Dlangle v_n^2 \Drangle$. 
The strongest non-linear response stems from the interactions
between $v_2$ and the other harmonics, and consequently a prominent response coefficient
is $w_{5(23)}/\e_2 \e_3$, which determines the $v_5$ produced by an
elliptic and triangular deformation.   
In \Sect{Hydroresponse} we determine these response
coefficients using both ideal and viscous hydrodynamics, and study how the
response depends on the shear viscosity. With these non-linear coefficients,
together with the linear response, we make several predictions for $v_1$,
$v_4$, and $v_5$ in ideal and viscous hydrodynamics in \Sect{results}.
Finally, in \Sect{results} we also study the transverse momentum dependence of
$v_1$, $v_4$, and $v_5$.  

In this work we will determine the harmonic spectrum by 
characterizing the quadratic response
of the system to  small deformations.
Alternatively, one could simply run hydrodynamics event-by-event
and compute the averages that are needed to compare to experiment \cite{Qiu:2011iv,Schenke:2010rr,Petersen:2010cw, Gardim:2011qn, Holopainen:2010gz, Werner:2010aa}. 
While  event-by-event hydrodynamics is the best for this
pragmatic purpose, the framework of non-linear response can yield valuable
insight into the physics of these rather involved simulations.

\section{Non-linear response}
\label{nonlin}

\subsection{The cumulant expansion}
In hydrodynamic simulations of heavy ion collisions the medium 
is first modeled with an initial state Glauber model,
then is evolved with hydrodynamics, and finally the particle spectrum
is computed by making kinetic assumptions about the fluid. 
The final state particle spectrum for each event  can be expanded in harmonics
\st
\label{spectra}
 \frac{dN}{ d\phi_\p  } 
= \frac{N}{2\pi}\,
\left(1 + \sum_{n=1}^{\infty} 
 v_n e^{i n (\phi_p - \Psi_n) } 
+ \mbox{c.c.} \right) \, ,
\stp
where here and below ${\rm c.c.}$ denotes complex conjugation.
The root mean squares of $v_n$ are easily determined experimentally,
and  are given a special notation
\st
   v_{n}\{2\} \equiv \sqrt{ \Dlangle v_n^2 \Drangle }  \, , 
\stp
where $\Dlangle \ldots \Drangle$ denotes the average over 
events.

In the next sections we will describe how the momentum space 
response is related to the initial state geometry.  
To this end, the spatial distribution of the initial entropy density in
the transverse plane, 
\st
\rho(\x)   \equiv \frac{\tau_0 s(\x)}{\int \dd^2x\, \tau_0 s(\x)} \, ,
\stp 
is quantified  with a cumulant expansion \cite{Teaney:2010vd},
where $\x = (x,y)=(r\cos\phi,r\sin\phi)$ are the coordinates in the transverse 
plane and $\tau_o$ is the  initial Bjorken time \cite{Ollitrault:1992bk}. 
Specifically  the $n,m$-th moment of the entropy distribution is  defined
as
\st
\label{momentdef}
  \rho_{n,m} \equiv \int \dd^2\x \,\rho(\x) \, (r^2)^{(n-m)/2} r^m e^{im\phi } 
  \, , 
\stp 
where $(n-m)/2$ is typically an integer.
This moment is closely related to the  $n,m$-th cumulant $W_{n,m}$ 
\st
\label{cumulant}
W_{n,m} \propto \rho_{n,m} - {\rm contractions}  \, .
\stp
The meaning of \Eq{cumulant} will be clarified through examples, with additional 
details about the cumulant expansion relegated to the literature \cite{Teaney:2010vd,Gubser:2010ui}.  
The radial variation of $\rho(\x)$
is quantified by the radial cumulants,  
$\sllangle r^2 \srrangle$ and 
$\sllangle r^4 \srrangle - 2 \sllangle r^2 \srrangle^2$,
while the 
the azimuthal  variation of $\rho(\x)$ is quantified by
the azimuthal cumulants
\begin{align}
 \e_1 e^{i \psie_1}  =&   -\frac{\sllangle r^3 e^{i \phis} \srrangle }{ \sllangle r^3 \srrangle   }  \, , \\
 \e_2 e^{i 2\psie_2}  =& -\frac{\sllangle r^2 e^{i 2\phis} \srrangle }{ \sllangle r^2 \srrangle } \, , \\
 \e_3 e^{i3\psie_3}  =& -\frac{\sllangle r^3 e^{i 3\phis} \srrangle }{ \sllangle r^3 \srrangle  } \, . 
\end{align}
Here $\llangle \ldots \rrangle$ denote an average over $\rho(\x)$ for a single event, and 
$\psie_1$,  $\psie_2$ and $\psie_3$ are the participant plane angles.
These coordinate space angles are distinct from the momentum space angles $\Psi_1,\Psi_2,$ and $\Psi_3$.

For the lowest harmonics the azimuthal cumulants and the azimuthal moments coincide,
and these definitions will appear obvious to most readers.
For the fourth harmonic  and higher,
we will  depart from traditional moment based definition, 
and quantify the deformations with cumulants rather than moments\footnote
{For $n\ge4$ we notate the cumulant based eccentricity by $\c_n$ to differentiate this quantity from the  
   moment based eccentricity $\epsilon_n$. $\c_n$ is equal to $W_{n,n}$ up to
   normalization and an overall factor of $\llangle r^n \rrangle$. 
}
\st
 \c_4 e^{i 4\psie_4 }  \equiv - \frac{1}{\sllangle r^4 \srrangle } \left[   \sllangle r^4 e^{i4\phis} \srrangle   - 3 \sllangle r^2 e^{i2\phis  } \srrangle^2  \right] \, .
\stp
The motivation for this definition can be seen by studying 
an elliptic Gaussian distribution,
\st
\label{background}
   \rho(\x) =  \frac{1}{2\pi \sigma_x \sigma_y}   e^{- \frac{x^2}{2 \sigma_x^2 }  - \frac{y^2}{2\sigma_y^2} } \,  , 
\stp
which has $\c_4 = 0$, although $\epsilon_4$ is non-zero 
and is of order  $\epsilon_2^2$.
%
Similarly we define
\st
 \c_5 e^{i5 \psie_5 }  \equiv  - \frac{1}{\sllangle r^5 \srrangle } \left[ 
\sllangle r^5 e^{i5\phis } \srrangle  
 - 10 \sllangle r^2 e^{i2\phis} \srrangle \sllangle r^3 e^{i3\phis} \srrangle \right] \, .
\stp
and remark that a Gaussian distribution deformed by an $\epsilon_3$,
\st
\label{triangle_space}
s(\x,\tau)   \propto   \left[1 +  \frac{\llangle r^3 \rrangle \epsilon_3}{24}  \left( \left(\frac{\partial }{\partial x} \right)^3 - 3 \left(\frac{\partial}{\partial y}  \right)^2 \frac{\partial}{\partial x} \right) \right] e^{- \frac{x^2}{2\sigma_x^2 } - \frac{y^2}{2\sigma_y^2 }  }
 \, , 
\stp
has $\c_5 = 0$, although $\sllangle r^5 e^{i5\phi_s} \srrangle$ is non-zero and of order 
$\epsilon_2\epsilon_3$.

We will characterize the hydrodynamic response to the cumulants defined above
in the next section.

\subsection{Non-linear response to the cumulants}
\label{vis_dep}

We expect the response of the system to be dominated by the lowest cumulants.
Motivated by Fourier analysis \cite{Teaney:2010vd}, we replace the general
distribution $\rho(\x)$ with a Gaussian, \Eq{background}, whose second moments
have been adjusted to reproduce $\llangle r^2 \rrangle = \sigma_x^2 +
\sigma_y^2$ and $\e_2 = (\sigma_y^2 - \sigma_x^2)/(\sigma_x^2 + \sigma_y^2)$.  
In \Ref{Teaney:2010vd} we showed that a Gaussian + fourth order cumulants
reproduces the results of smooth Glauber initial conditions in  detail.
If a Gaussian with a non-negligible $\epsilon_2$ is simulated, the 
particle spectrum produced by this background contains all even harmonics
\st
\label{w422}
 \frac{dN}{d\phi_\p}  = \frac{N}{2\pi}\, \big(1 +   w_2 e^{i2(\phi_\p - \psie_2)
} +   w_{4(22)} e^{i{4(\phi_\p - \psie_2)} }  + \ldots + \mbox{c.c.} \big) \, .
\stp
For small $\e_2$  the response coefficient 
$w_2$ describes the linear response to the deformation and is proportional to $\e_2$,
while $w_{4(22)}$ describes  the non-linear response and is proportional to $\e_2^2$. 
Below, we will assume that $\e_2$ is small enough that this scaling with $\e_2$ 
applies. Further, we have truncated the expansion in \Eq{w422} at quadratic
order in $\e_2$, and will continue to do this implicitly from now on. 
The working assumption in this paper is that the most important
non-linearity stems from the almond shape of the background.


If the Gaussian distribution is perturbed by 
a small fourth order cumulant $\c_4 e^{i4\psie_4}$,
then the  resulting particle spectra will be  described  by
\st
\label{w4w422}
 \frac{dN}{d\phi_\p}  =\frac{N}{2\pi}\left(1 +  w_2 e^{i2(\phi_\p - \psie_2) } +   w_4 e^{i 4(\phi_\p - \psie_4) }   +   w_{4(22)} e^{i{4(\phi_\p - \psie_2)} } +   \mbox{c.c.} \right) \, , 
\stp
where $w_4$ captures the linear response to the fourth order cumulant and
is proportional to $\c_4$ for small $\c_4$.
In writing  \Eq{w4w422} we have neglected terms proportional to
$\c_4\epsilon_2$,  which can contribute to $v_2$ 
and  reduce the perfect correlation  between $\Psi_2$ and $\Phi_2$.
Comparing \Eq{w4w422} with the definition of $v_4$, \Eq{spectra}, we see that
$v_4$  is determined by the linear and 
 quadratic response 
\st
 v_4 e^{-i 4\Psi_4 }  =  w_4 e^{-i4\psie_4}  +  w_{4(22)} e^{-i4\psie_2}  \, .
\stp
Squaring this result and averaging over events we see that  
\st
\label{v4sq}
 v_{4}\{2\} \equiv \Dlangle v_4^2 \Drangle^{1/2}  =  \Dlangle |  w_4 e^{-i4\psie_4 } +  w_{4(22)} e^{-i4 \psie_2 } |^2 \Drangle^{1/2} \, .
\stp
In  writing \Eq{w4w422} we have neglected the  non-linear contributions
of $\e_1$ and $\e_3$ to $v_4$ since $v_3$ and $v_1$ are small compared
to $v_2$ for mid-peripheral collisions. 

Similarly, if the Gaussian background distribution is perturbed by 
a third order cumulant and a fifth order cumulant $\c_5$,
then $v_5$ is determined by a combination of the linear and non-linear response.
The response to $\c_5$ is small \cite{Alver:2010dn}, and therefore we will neglect the non-linearities 
due to $\e_2 \c_5$, but we will keep the non-linearities due to $\e_2 \e_3$.
With this approximation  scheme the particle spectrum through quadratic order reads 
\begin{multline}
\label{dndp_23}
 \frac{dN}{d\phi_\p} = \frac{N}{2\pi}\, \Big(1  +   
w_3 e^{i3(\phi_\p - \psie_3) }  + 
w_5 e^{i5(\phi_\p - \psie_5) }  +     \\
w_{1(23) } e^{i \phi_\p - 3\psie_3 + 2\psie_2 }  +       
w_{5(23) } e^{i 5\phi_\p - 3\psie_3 - 2\psie_2}  + \mbox{even harmonics} + {\rm c.c } \Big) \, .
\end{multline}

Comparing this equation  to the definition of $v_5$, we see that
\st
\label{v5sq}
   v_5\{2\} = 
\Dlangle | w_5 e^{-i5\psie_5} \, +  \, \, w_{5(23)} e^{-i(3 \psie_3 + 2 \psie_2)  }|^2 \Drangle^{1/2} \, ,
\stp
which is clearly analogous with $v_4$ case.
Finally,  
if the distribution has a net dipole asymmetry $\epsilon_1$,
then $v_1$  is given a combination of the linear and non-linear response
\st
\label{v1sq}
  v_1\{2\}  =  \Dlangle |  w_1 e^{-i\psie_1 }  +  \, w_{1(23)} e^{-i (3\psie_3 -2\psie_2)  } |^2 \Drangle^{1/2} \, ,
\stp
where $w_1$ notates  the linear response to $\epsilon_1$.  In writing this
result for $v_1$ we have neglected the non-linear  interaction between 
$v_1$ and $v_2$, {\it i.e.} $w_{1(21)}$. Thus \Eq{v1sq} makes the simplifying
assumption that $v_1$ is small compared to $v_3$, while a more complete treatment would include a $w_{1(21)}$ contribution.
 
Let us discuss how this formalism can be used to study the $p_T$ dependence of the flow. The particle spectra is
expanded in harmonics 
\st
   \frac{\dd N}{\dd p_T \dd \phi_\p}  \equiv \frac{\dd N}{\dd p_T} \left(1 + \sum_{n=1}^{\infty}  v_n(p_T) e^{in(\phi_\p -\Psi_n(p_T)) }  + {\rm c.c.} \right)  \, ,
\stp
where the phase, $\Psi_n(p_T)$,  is in general a function of $p_T$. Then $v_n(p_T) \{2 \}$ in the $\Psi_n$ plane is normally defined as 
\st
\label{vnptdef}
v_n(p_T)\{2\} \equiv  \begin{cases}
     \frac{ \Dlangle v_n(p_T) v_n \cos(n(\Psi_n(p_T)-\Psi_n) ) \Drangle }{ v_n\{2\} }  
    &  n > 1 \\
     -\frac{ \Dlangle v_1(p_T) v_1 \cos(\Psi_1(p_T)-\Psi_1) \Drangle }{ v_1\{2\} }  
    & n = 1 
\end{cases} \, , 
\stp
where we have inserted an extra minus sign 
for $v_1(p_T)$,  since the integrated $v_1$ is negative.  The phase
angle $\Psi_n(p_T)$ is often assumed to equal $\Psi_n$.
Using the formalism outlined above  we write  $v_1(p_T)$ as a sum of the linear and non-linear response
\st
v_1(p_T) e^{-i\Psi_1(p_T)} = w_1(p_T) e^{-i \psie_1} +  w_{1(23)}(p_T) e^{-i3\psie_3
+  i 2\psie_2} \, .
\stp
Then the numerator of $v_{1}(p_T) \{2\}$ is given by
\begin{multline}
   \label{numeratorv1}
     \Dlangle v_1(p_T) v_1 \cos(\Psi_1(p_T) - \Psi_1) \Drangle =  \\
     \Bigdlangle 
  w_1(p_T) w_1 + w_{1(23)}(p_T) w_{1(23)} + \left[ w_{1}(p_T) w_{1(23)} + w_{1(23)}(p_T) w_{1} \right] \cos(\psie_1 - 3\psie_3 + 2\psie_2)
\Bigdrangle  \, ,
\end{multline}
and the denominator is given by the integrated expression for $v_{1}\{2\}$, \Eq{v1sq}.  Similar expressions follow for $v_4(p_T)$ and $v_5(p_T)$.

Finally, let us place some older measurements and calculations of $v_4(p_T)$
into context \cite{Kolb:2003zi,Borghini:2005kd,Gombeaud:2009ye,Luzum:2010ae,Luzum:2010ad,Bai:2007ky,Adare:2010ux}. Traditionally, what was referred to as $v_4(p_T)$ would today be called $v_4(p_T)$ in the $\Psi_2$ plane:
\st
v_{4(22)}(p_T)\{2\} \equiv  \frac{\Dlangle v_4(p_T) v_2 \cos(4\Psi_4(p_T) - 2\Psi_2 - 2\Psi_2) \Drangle }{ v_2\{2\} } \, .
\stp
As discussed in the conclusions, the differences between $v_{4(22)}(p_T)\{2\}$ and $v_4(p_T)\{2\}$ can be used to partially disentangle the linear and non-linear  response.

\subsection{Summary}

The goal of the present work is to compute the linear and non-linear response coefficients, and to use these coefficients together with an initial 
Glauber model to determine $\Dlangle v_n^2 \Drangle$ 
with  Eqs.~(\ref{v4sq}),(\ref{v5sq}), and (\ref{v1sq}).  
For $v_5$   the step by step procedure is:\, 
 (i) use hydrodynamics to determine
the response coefficients
\st
     \frac{w_5}{\c_5}\, , \qquad \mbox{and}   \qquad   \frac{w_{5(23)}}{\e_2\e_3}\,  ,
\stp
for vanishingly small $\c_5$ and $\e_2\e_3$; \, (ii)
use a Glauber model to determine the geometric coefficients that are needed in \Eq{v5sq}, 
$\Dlangle \c_5^2 \Drangle$, $\Dlangle (\e_2\e_3)^2 \Drangle$,
and $\Dlangle \c_5 \e_2 \e_3 \cos(5\psie_5 - 3\psie_3 - 2\psie_2) \Drangle$; \, 
(iii) combine these results in \Eq{v5sq} to determine the complete hydrodynamic
prediction for $\Dlangle v_5^2 \Drangle$. 
The necessary Glauber correlations are determined using the Phobos 
Monte Carlo Glauber Model \cite{Alver:2008aq}, and
we note  that there 
is  a very strong geometric correlation  between participant planes
differing by  two, {\it e.g.}
\st
   \Dlangle \c_5 \e_2 \e_3  e^{i(5\psie_5 - 3\psie_3 - 2 \psie_2) }  \Drangle\, ,  
\qquad  \mbox{and} \qquad 
   \Dlangle \e_1 \e_2 \e_3  e^{i(3\psie_3 - \psie_1  - 2 \psie_2) } \Drangle \, . 
\stp
This geometric correlation can be 
studied analytically in an independent source model \cite{Bhalerao:2011bp}, and is 
easily attributed to the elliptic shape of the 
overlap region \cite{Teaney:2010vd,Bhalerao:2011bp,Jia:2012ju}.

\section{Hydrodynamic Simulations} 

\subsection{Ideal and viscous hydrodynamics}
\label{hydro_intro}

To calculate the non-linear response we use a hydrodynamics code that implements conformal second order hydrodynamics \cite{Baier:2007ix}. 
The numerical scheme is based  on a central scheme developed and tested in \Ref{Dusling:2007gi},
although the equations of motion for the $\pi^{ij}$ are somewhat different from
what was studied in that work\footnote{ 
However, when additional non-conformal second order gradients  are added to our equations of motion and the parameters are matched, our current numerical can be compared directly to \Ref{Dusling:2007gi}.
If this is done, the two hydro-codes yield the same answers to $0.1\%$
for the type of problems considered in this work.
}.
$\eta/s$ is held constant, and the ratio of second order hydro parameters are
taken from their AdS/CFT values \cite{Baier:2007ix,Bhattacharyya:2008jc},
{\it e.g.} $\tau_\pi/(\eta/sT) =4-2\ln2$.
The equation of state partially parametrizes lattice results 
and was  used previously by Romatschke and Luzum \cite{Luzum:2008cw}.
Finally, we have followed the time ``honored" constant temperature freezeout
prescription, with $T_{\rm fo} = 150\,{\rm MeV}$. For simplicity we have adopted  the popular quadratic ansatz for the viscous correction to the thermal distribution 
function \cite{Teaney:2009qa}
\st
\label{deltaf}
f(P) = f_o(P)  + \delta f(P) \, ,    \qquad \delta f(P)  \equiv \frac{f_o (1 \pm f_o)}{2(e + \mathcal P) T^2} P^{\mu} P^{\nu} \pi_{\mu\nu} \,  ,
\stp
where $f_o(P) = 1/(\exp(-P\cdot U(X)/T)  \mp 1)$ is the equilibrium 
distribution, $e + \mathcal P$ is the enthalpy,  and $\delta f$ is the first viscous correction \cite{Teaney:2003kp,Teaney:2009qa}.   
Although we have used the quadratic ansatz in this work,  a linear ansatz is probably more appropriate for QCD-like theories and can effect the integrated 
flow for the higher harmonics \cite{Dusling:2009df,Luzum:2010ae}. 

For the simulations shown below we have followed the centrality classification 
given in \Ref{Qiu:2011iv} which is documented in Table I. of that work.    
Our procedure to determine the response coefficient at a given impact
parameter largely follows \Ref{Teaney:2010vd}, which should be referred 
to for additional details -- see especially Appendix A of that work. 
Briefly, 
for each impact parameter we determine the average squared radius $\llangle r^2 \rrangle$, and
 initialize a Gaussian distribution that is deformed  
by the appropriate cumulant.  The Gaussian is normalized to reproduce the total entropy
in the event.
For instance, to determine the $w_{5(23)}$ we initialize the distribution 
given in \Eq{triangle_space} with $\epsilon_2=\epsilon_3=0.02$.  
A technical complication is that  the distribution in \Eq{triangle_space} must be regulated \cite{Teaney:2010vd}, and the regularization  procedure introduces a small $\c_5$. However, 
the spurious $\c_5$ decreases faster than $\epsilon^3$  and can be made 
arbitrarily small  compared to the signal. 
Empirically we find that
the spurious $\c_5$ decreases approximately as $\epsilon^5$, and the 
$v_5$ from the spurious cumulant is negligibly small 
compared to the $v_5$ from the $\epsilon_2\epsilon_3$ combination. 

\subsection{The non-linear response coefficients in ideal  and viscous hydrodynamics} 
\label{Hydroresponse}

In this section we will study  the non-linear response coefficients  systematically. 
In particular we study 
how the linear and non-linear response coefficients depend
\begin{inparaenum}[(i)]
\item  transverse momentum,
\item centrality, 
and \item  shear viscosity. 
\end{inparaenum} 

\subsubsection{Momentum dependence of the response coefficients }

\begin{figure}
\includegraphics[width=0.49\textwidth]{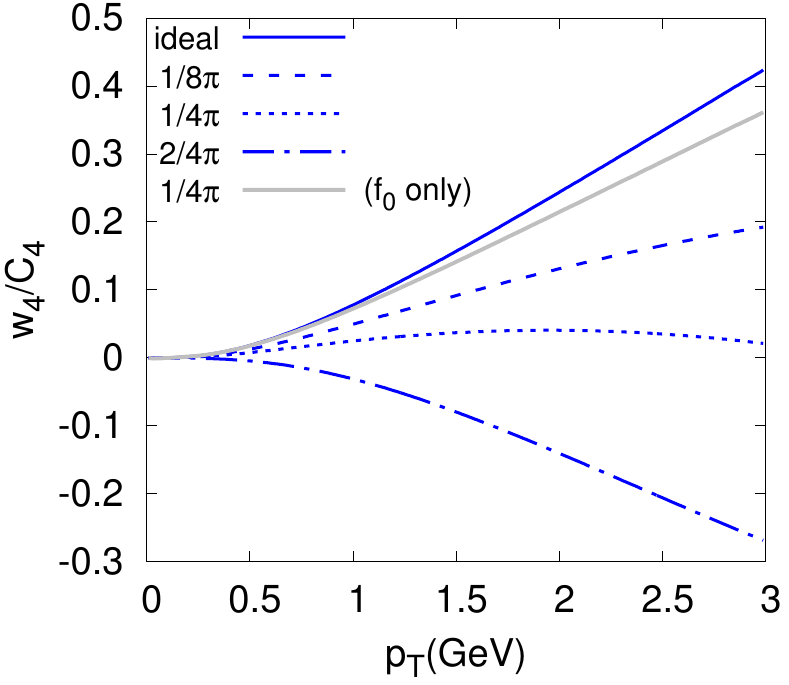}
\includegraphics[width=0.49\textwidth]{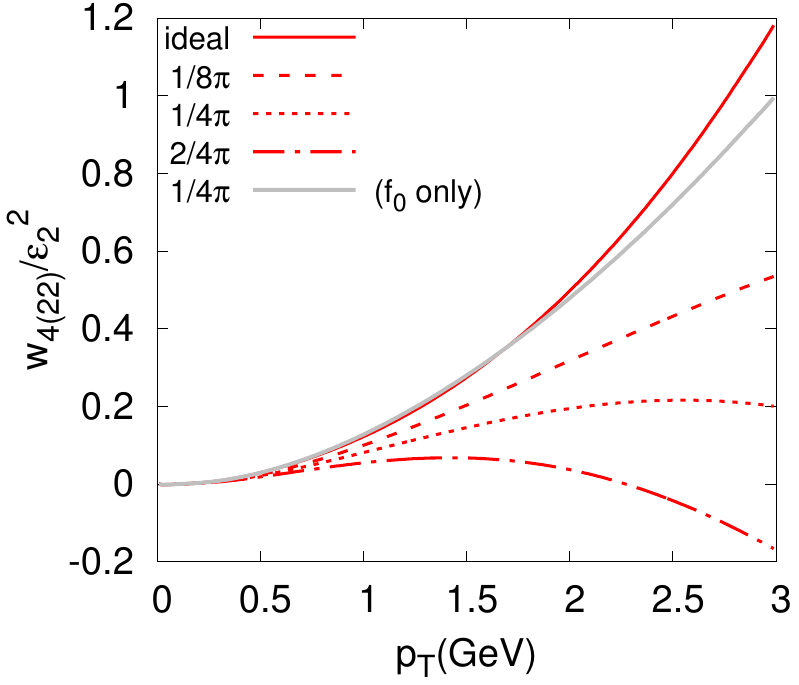}
\caption{
   (color online) The linear and non-linear response coefficients for $v_4$, $w_4(p_T)$ and $w_{4(22)}(p_T)$,
   for ideal and viscous hydrodynamics. The grey curves (which are 
   shown only for $\eta/s=1/4\pi$)  exhibit the 
   resulting response when the equilibrium distribution  $f_o$ is used, and the viscous correction, $\delta f$, is neglected (see \Eq{deltaf}).
\label{wnresponse}
 } 
\end{figure}

\Fig{wnresponse} examines the $p_T$ dependence of the linear
and non-linear response coefficients, $w_4$ and $w_{4(22)}$, which  
are characteristic of the response coefficients more generally. First, focus on the ideal curves in
\Fig{wnresponse}(a) and (b). 
At large $p_T$ the non-linear response curves 
show a characteristic quadratic rise with $p_T$, while the linear response 
curves show a characteristic  linear rise.  This
difference between the non-linear and linear response is known from previous
studies of $v_4$ \cite{Borghini:2005kd}. Later, when examining non-linear
corrections to $v_1$ (see \Fig{v1scaling_pt}), we will see that the non-linear corrections are most important at high $p_T$ and exhibit a characteristic quadratic rise.
Comparing \Fig{wnresponse}(a) and (b),  we see that  
viscous corrections are smaller for  the non-linear response
$w_{4(22)}(p_T)/\epsilon_2^2$, than for the linear response $w_4(p_T)/\c_4$.
This is a generic result as will be discussed in detail in \Sect{viscdependence}.

We also note that the linear response curves shown in \Fig{wnresponse}(a) change sign for sufficiently large viscosity. This is an artifact of the first viscous correction, $\delta f$, and  the quadratic ansatz. To see this, we have plotted  $w_4(p_T)$ and 
and $w_{4(22)}(p_T)$ using only the unmodified  distribution function $f_o$ in
\Fig{wnresponse}(a) and (b).
For large viscosity the $\delta f$  correction to $v_4$ and $v_5$  
is not small compared to the ideal contribution $f_o$, and this causes a  reduction 
of the response, which is more pronounced for the higher harmonics, $v_4$ and $v_5$.
In full kinetic theory calculations $w_4/\c_4$ and $w_5/\c_5$ remain positive 
and approach zero as the viscosity is increased \cite{Alver:2010dn}. Thus, the negative
$w_4/\c_4$ indicates that the first viscous correction has become too large to
be trusted.  Below, we will simply set the response coefficients  to zero when
this is the case. Experience with kinetic theory suggests that this ad hoc procedure
is not far from what really happens.

\subsubsection{Centrality dependence of the response coefficients}
\begin{figure}
\includegraphics[width=0.49\textwidth]{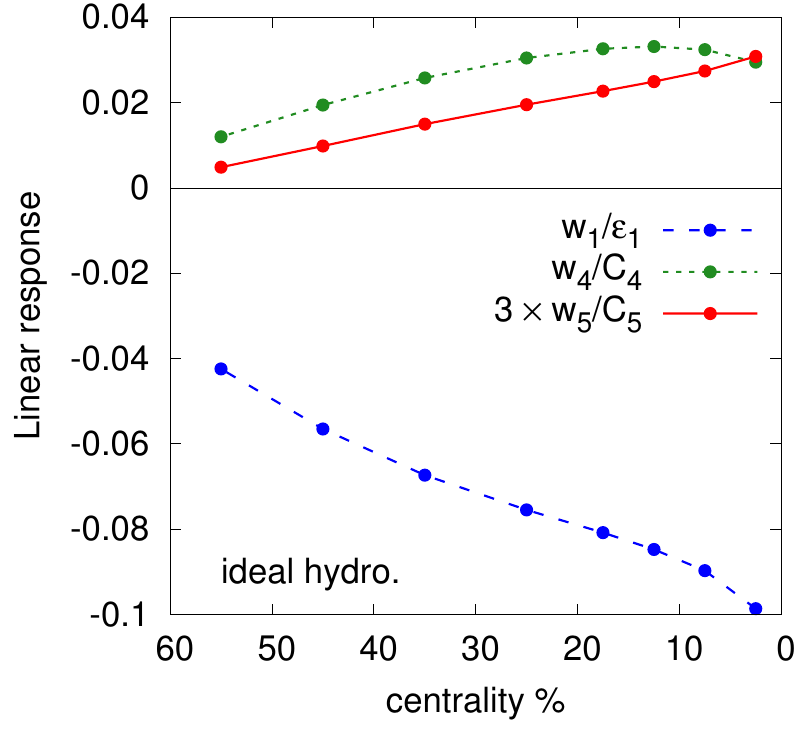}
\includegraphics[width=0.49\textwidth]{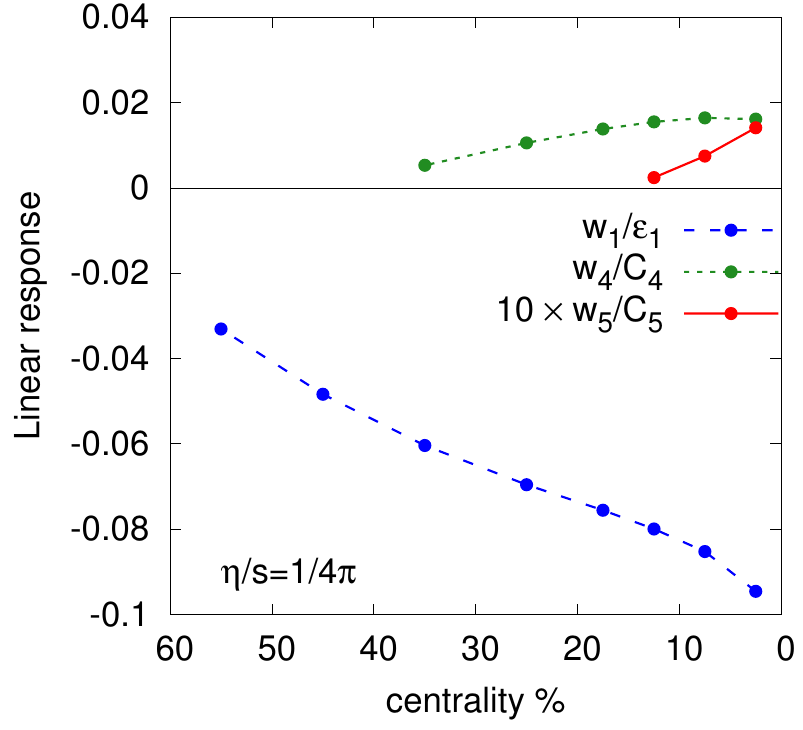}
\includegraphics[width=0.49\textwidth]{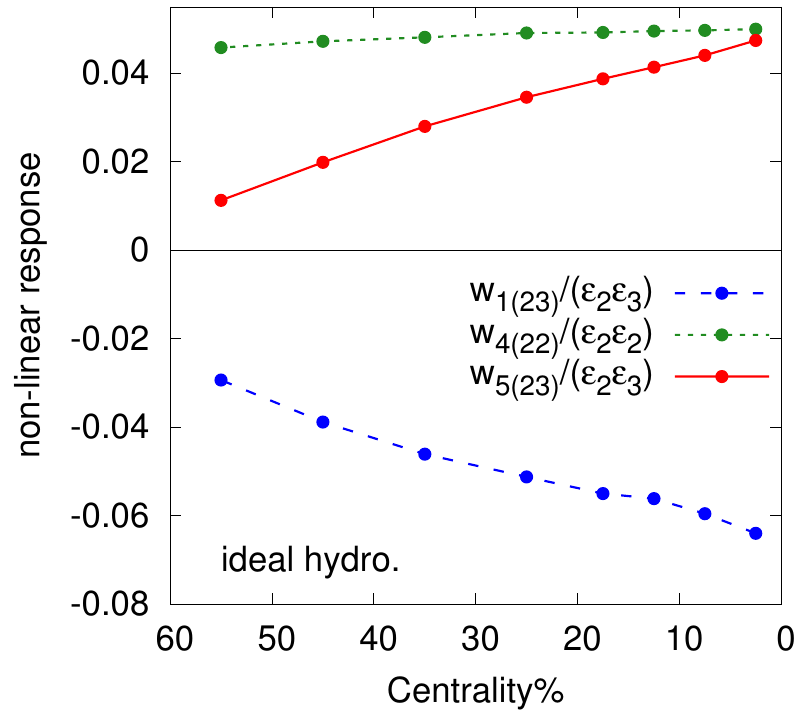}
\includegraphics[width=0.49\textwidth]{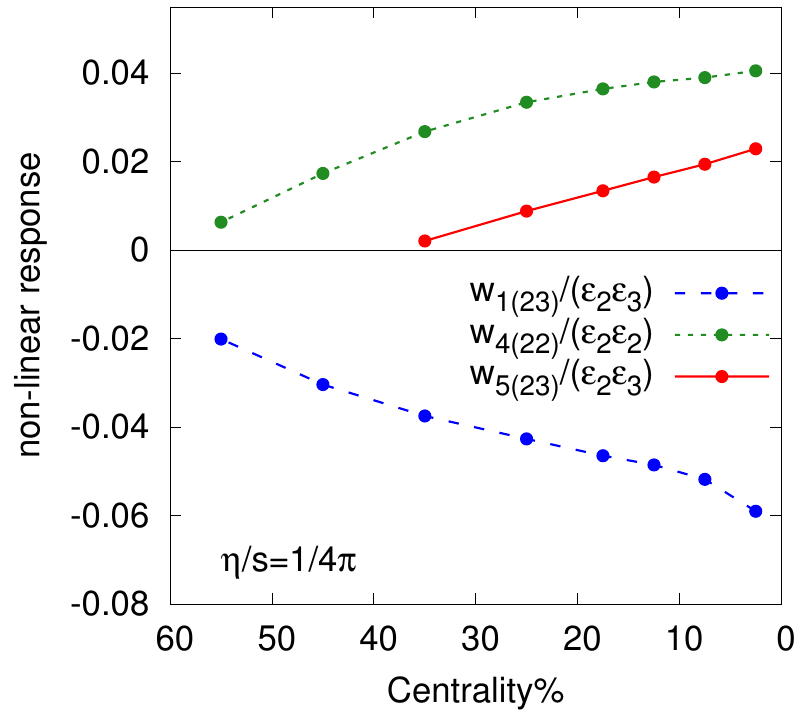}
\caption{
The linear and non-linear response coefficients for ideal and viscous hydro.
In the viscous case the curves are truncated when the response coefficients turn
negative, {\it i.e.} outside of the regime of validity of viscous hydro.
\label{response_fig}
 } 
\end{figure}

\Fig{response_fig}  shows  the linear
and non-linear response coefficients in ideal and viscous hydrodynamics.
There are several salient features  contained in these plots.
First, note that the magnitude of the linear response coefficient
$w_5/\c_5$ is quite small in the viscous case, and $w_5/\c_5$ has
been multiplied by ten to make the curves visible. The non-linear response
$w_{5(23)}$  coefficient is significantly larger. The implications
of this difference will be studied in the next section when we multiply
the response coefficients by 
$\c_5$ and $\epsilon_2 \epsilon_3$ 
respectively.
Second, all of the response coefficients are reduced by viscosity, 
especially in non-central collisions.

The viscous $w_4/\c_4$ and $w_5/\c_5$ curves stop abruptly as a function of centrality, since
we have truncated the curves when response falls below zero.  
As discussed above (see \Fig{wnresponse}), this is because  viscous
corrections to the thermal distribution function ($\delta f$) become larger for more
peripheral collisions, and
this correction is magnified by the high harmonic number.
We have therefore truncated the $w_4$ and $w_5$ response curves when the response turns negative.
At this point $\delta f$ constitutes an order one correction and can no longer be trusted.

\subsubsection{Dependence on viscosity}
\label{viscdependence}
\begin{figure}
\includegraphics[width=0.49\textwidth] {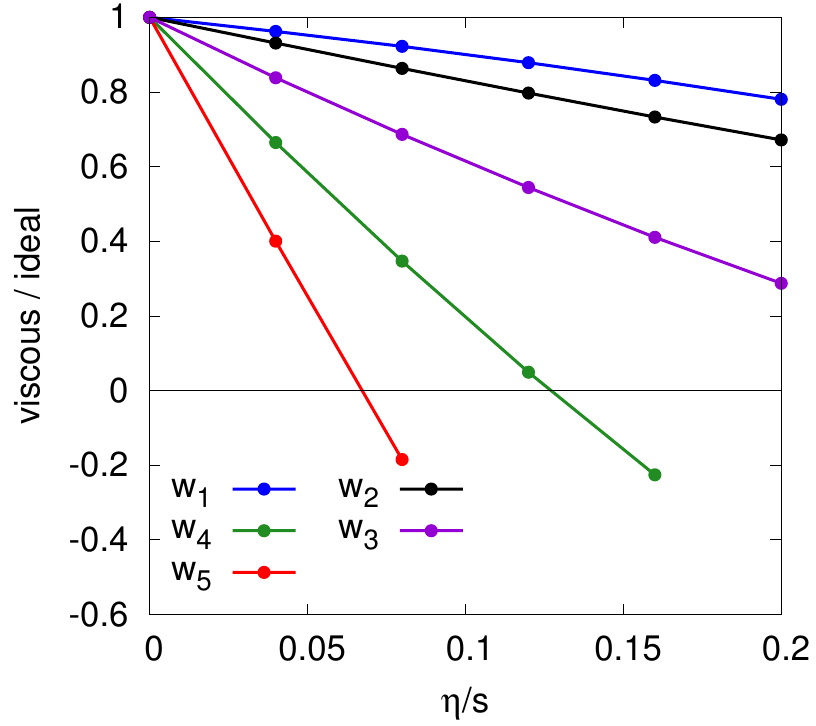}
\includegraphics[width=0.49\textwidth] {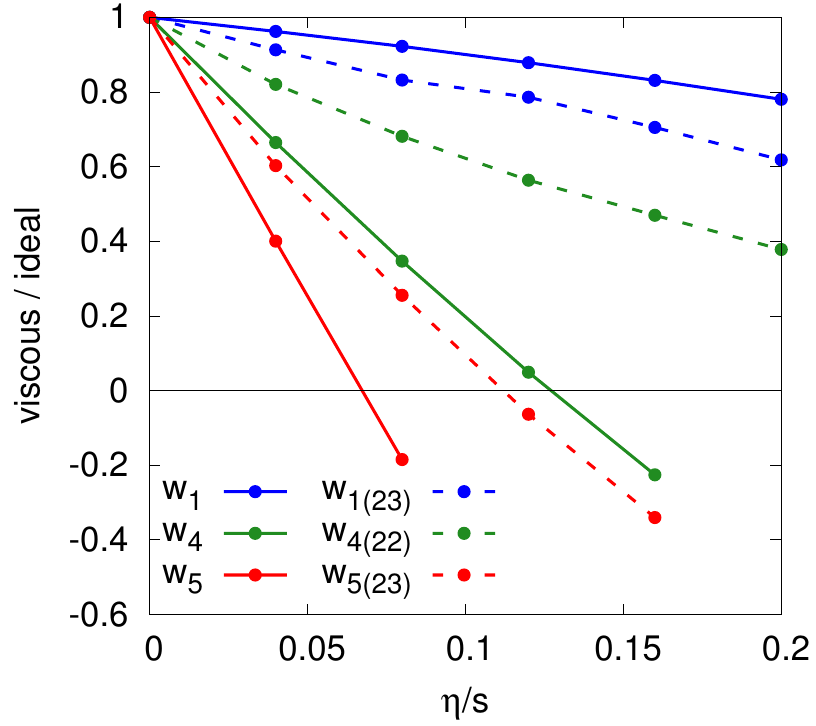}
\caption{
\label{visc_response}
(a) Linear response coefficients $w_n$ as 
a function of viscosity relative to the ideal response. 
(b) A comparison of non-linear and linear response coefficients 
as a function of viscosity.  {\it e.g.} $w_{5(23)}$ 
records the $v_5$ produces by a combination of $\epsilon_2$ 
and $\epsilon_3$.  The negative values for large viscosity are
spurious, and lie beyond the region of applicability of viscous hydrodynamics.
}
\end{figure}

It is interesting to note that viscous reduction for $w_1/\epsilon_1$ is smaller than for $w_4/\c_4$ and $w_5/\c_5$.  
This pattern of viscous corrections for linearized perturbations is studied further in \Fig{visc_response}(a). 
Each linearized perturbation  labeled by $n,m$-th cumulant is damped by a factor $\sim \exp(-\Gamma_{n,m}\, \Tf)$ relative to ideal hydrodynamics, where $\Tf$ is an estimate
for the duration of the event.   
Analytical work shows that
the damping  coefficients $\Gamma_{n,m}$ scale as
\st
\label{linscale}
   \Gamma_{n,m} \, \Tf \sim  \frac{\ell_{\rm mfp}}{L} \left( \frac{n-m}{2} + m \right)^2 \, ,
\stp
for a conformal equation of state and  a particular background flow \cite{Gubser:2010ui}.  
Thus,  each power of $r^2$ and each harmonic order  in \Eq{momentdef} increases
$(n{-}m)/2{+}m$ by one unit.  Our numerical work (\Fig{visc_response}(a)) is not limited to the conformal equation of state 
or the particular background flow of \Ref{Gubser:2010ui}, and shows that this scaling is
reasonably generic \cite{Alver:2010dn,Retinskaya:2012ky}.  Specifically, the formal estimate given in \Eq{linscale} implies a definite pattern among the viscous corrections to $v_n$:
\st
   -\frac{\Delta w_1}{w_1^{\rm id}} \simeq -\frac{\Delta w_2}{w_2^{\rm id}}   \propto \, 4 \, \frac{\eta}{s} \, ,  
   \qquad 
-\frac{\Delta w_3}{w_3^{\rm id}}  \propto \, 9 \, \frac{\eta}{s} \, ,  
   \qquad -\frac{\Delta w_4}{w_4^{\rm id}}  \propto \, 16 \, \frac{\eta}{s}
\, ,   \qquad -\frac{\Delta w_5}{w_5^{\rm id}}  \propto \, 25 \, \frac{\eta}{s} \, .
\stp
where $\Delta w=w^{\rm viscous} -  w^{\rm ideal}$, and $w^{\rm id}$ is
the ideal hydro response coefficient.
Note, in particular that the viscous corrections $v_1$ and $v_2$ 
are similar since $v_1$ and $v_2$ respond to the
dipole asymmetry, $W_{3,1}$, and the ellipticity, $W_{2,2}$, respectively \cite{Retinskaya:2012ky}.
Since the slopes of the $v_1:v_2:v_3:v_4:v_5$ curves
in \Fig{visc_response}(a)  have approximately
the expected ratios $4:4:9:16:25$,  our numerical work
 qualitatively confirms this pattern of viscous corrections. 

\Fig{visc_response}(b) compares the damping rate
for the non-linear response  coefficients  to the corresponding linear response coefficients.
Take $w_{5(23)}$ for example. Since $w_{5(23)}$ is of order
$~ v_2 v_3 $ we expect the damping of this non-linear perturbation to
scale as $\sim e^{-\Gamma_{2,2} \tau} e^{-\Gamma_{3,3} \tau}$, and thus the
damping rate  $\Gamma_{5(23)}$ is expected to scale as
\st
\label{non-linear_damp}
    \Gamma_{5(23)}  \sim \Gamma_{2,2} + \Gamma_{3,3}  \, .
\stp
Thus, we expect  the non-linear  and linear
response coefficients for $v_5$ to scale as
\st
-\frac{\Delta w_{5(23) } } 
{w_{5(23)}^{\rm id} } \propto 13 \frac{\eta}{s}  \, ,
\qquad 
-\frac{\Delta w_{5} } 
{w_{5}^{\rm id} } \propto 25 \frac{\eta}{s}   \, .
\stp
Comparing the slopes of the non-linear and linear response curves in \Fig{visc_response}(b), we see
that the slope of the $\Delta w_{5(23)}/w_{5(23)}^{\rm id}$  curve is approximately half of the
corresponding $\Delta w_{5}/w_5^{\rm id}$,  and is qualitatively consistent
with our heuristic estimate of $13/25$.  $w_{4(22)}$ and $w_{4}$ 
show a similar pattern  of viscous corrections. 
Finally our estimates seem only partially 
applicable to $v_1$.
For instance, the reasoning of \Eq{non-linear_damp} predicts that
the non-linear damping rates,  $\Gamma_{1(23)}$ and $\Gamma_{5(23)}$,
should be  equal. However, 
the slope of $\Delta w_{1(23) }/w_{1(23)}^{\rm id}$ is 
significantly smaller than the $\Delta w_{5(23)}/w_{5(23)}^{\rm id}$,
and contradicts this reasoning.
Clearly, the non-linear viscous damping 
of  $v_1$ is a special case which  will have to be investigated more completely at a later date.

\section{Results and Discussion}
\label{results}

\subsection{Results}
\begin{figure}
\includegraphics[width=0.9\textwidth]{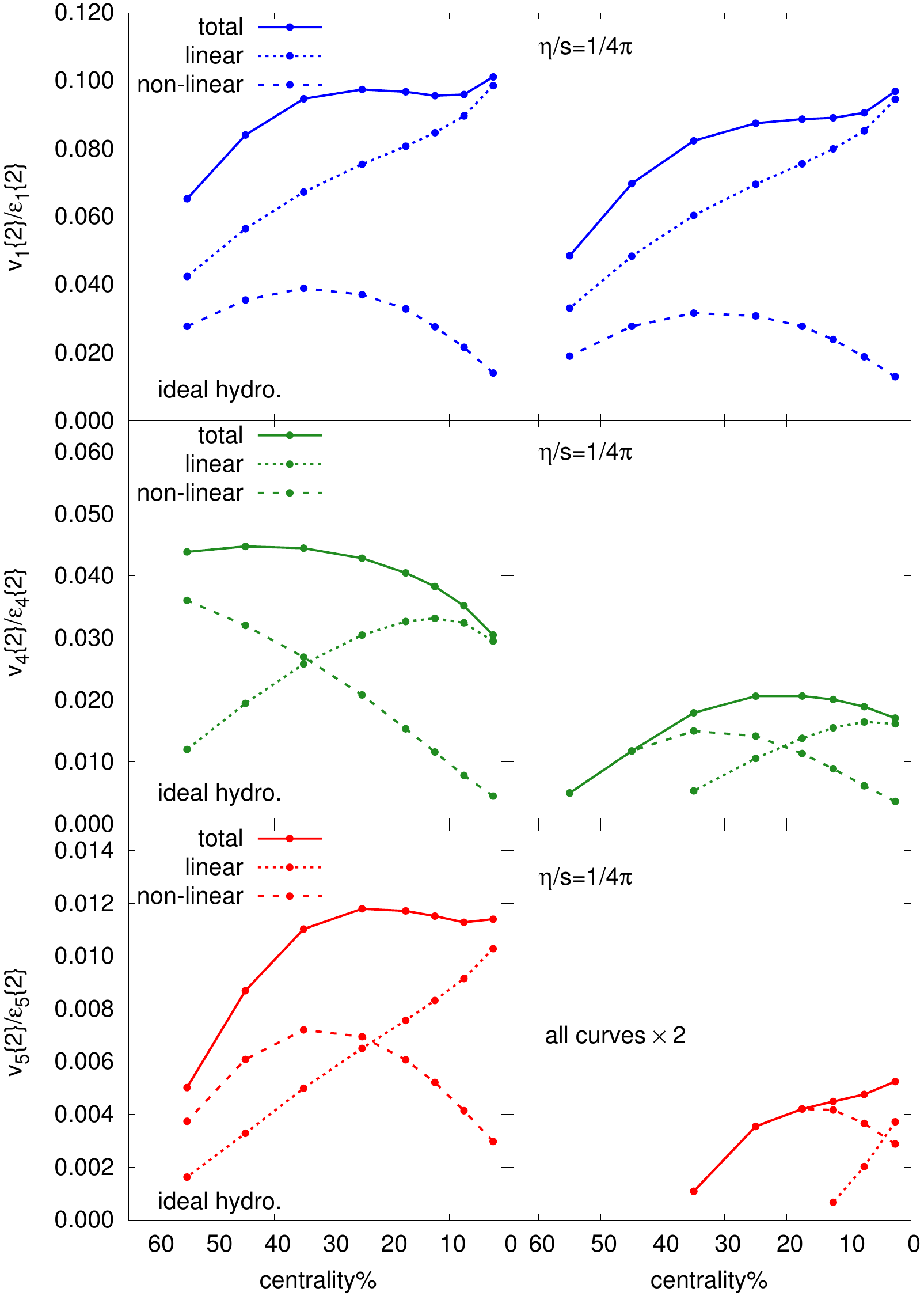}
\caption{
$v_1$, $v_4$ and $v_5$ versus centrality in ideal and viscous hydrodynamics.
To keep the ideal and viscous curves on the same scale we have multiplied
the viscous $v_5$ curves by a factor of two.
In the viscous case, the linear response is neglected when the 
response coefficients turn negative, {\it i.e.} outside of the
region of applicability of viscous hydrodynamics.
\label{vn_fig}
 } 
\end{figure}

Having clarified the non-linear hydrodynamic response, we study 
the phenomenological implications of these response coefficients.
\Fig{vn_fig} shows $v_1$,  $v_4$ and $v_5$ including the linear and non-linear response as outlined in \Sect{nonlin}, and is the principal result of this
work.   

Examining this figures we see that the non-linear response
is an important correction for $v_1$, and essential for $v_4$ and
$v_5$. The contribution of the non-linear response to the total flow increases
towards peripheral collisions,
and for $v_4$ and $v_5$ is of order $50\%$  in mid-peripheral collisions.
This is roughly compatible with simulation results from  event-by-event hydrodynamics \cite{Qiu:2011iv,Gardim:2011xv}.
Especially for viscous hydrodynamics and for $v_5$,
the linear response is negligible in all but the most central
bin. Even in the most central bin, the non-linear contribution
to $v_5$  is about  50\% of the total. It is notable, if expected, that 
for $v_1$ viscosity reduces the
non-linear contribution relative to the total, 
while for $v_4$ and $v_5$ viscosity
increases the non-linear contributions. This is consistent
with the discussion given in \Sect{vis_dep}.

It will be quite interesting to measure the 
complete set of event planes $(\Psi_1,\Psi_2,\Psi_3, \Psi_4, \Psi_5)$
and their inter-correlations. These measurements will place a strong 
experimental constraint on  the relative of importance 
of the non-linear response \cite{ATLASCorrelations}. For example, if the non-linear response is dominant (as implied by the viscous $v_5$ curves), then a 
stronger than geometric 
correlation is expected for certain experimental averages, {\it e.g.} $\llangle \cos(5\Psi_5 -3\Psi_3 -2\Psi_2) \rrangle$.

Next we examine the $p_T$ dependence of the $v_1$,$v_4$ and $v_5$.  Since $v_4$
and $v_5$ are  dominated by the non-linear response we will present our results by
scaling $v_2^2$ and $v_2v_3$ respectively. Many of the points
raised in this and the next paragraph are familiar from earlier  
studies of $v_4$ in the $\Psi_2$ plane.  In particular, the 
importance of non-linearities and fluctuations in determining 
the experimental $v_4/v_2^2$ ratio was understood previously \cite{Borghini:2005kd,Gombeaud:2009ye}.

First we note that according to an old argument by Borghini
and Ollitrault \cite{Borghini:2005kd},
$v_4/v_2^2$ should approach 
$1/2$  at large momentum in ideal hydrodynamics for any given event 
due to the non-linearities inherent in the phase space distribution.
Their result is easily generalized to $v_5$, $v_5 =v_2 v_3$. The argument 
follows by computing the freezeout distribution
in a saddle point approximation \cite{Blaizot:1986bh}, and can be schematically
understood by examining the thermal factor in 
an approximately radially symmetric flow profile. 
The transverse
flow vector as a function  of the spatial azimuthal  angle $\phi$ 
relative to the reaction plane is 
\st
\vec{u}_T = (u^x, u^y) \simeq (u_T(\phi) \cos\phi , u_T(\phi) \sin\phi) \, ,
\stp
where in the second step we have assumed that the flow is approximately
radially symmetric.
The transverse flow velocity  is then expanded in harmonics
\st
u_T(\phi) = u_{T}^{(0)} + 2 u_{T}^{(2)} \cos2\phi + 2 u_{T}^{(4)} \cos4\phi + \mbox{other harmonics}   \, ,
\stp
and the thermal factor  with $\vec{p} = (p_T\cos\phi_\p, p_T\sin\phi_\p)$ 
reads 
\begin{align}
e^{\vec{p}\cdot \vec{u}/T} 
\simeq& 
e^{\frac{p_T}T u_T^{(0)} \cos(\phi_\p - \phi)} \left[1 + \frac{2p_T}{T}
u_{T}^{(2)} \cos2\phi +  
\frac{1}{2} \left(\frac{2p_T}{T} u_{T}^{(2) } \cos2\phi \right)^2 + \ldots
\right] \, , \\
\simeq& 
e^{\frac{p_T}T u_T^{(0)} (\cos\phi_\p - \phi)} \left[1 +  \frac{2p_T}{T}
u_{T}^{(2)}\cos2\phi_\p +   
\left(\frac{p_T}{T} u_{T}^{(2)}\right)^2 
\cos4\phi_\p + \ldots  \right] \, .
\end{align}
The leading exponential strongly correlates coordinate space angle $\phi$ and 
the momentum space angle $\phi_\p$.
In the second line we have anticipated the saddle point  approximation,
(which realizes this correlation)
and set $\phi\simeq\phi_\p$ in the post-exponent.
The second term in square brackets
determines the linear response
coefficient $w_2$ and rises linearly with momentum, 
$w_2 \sim p_T u_T^{(2)}/T$.
The third term determines
 the non-linear response coefficient $w_{4(22)}$, and grows 
quadratically with momentum, $w_{4(22)} \sim \half  (p_T u_T^{(2)}/T)^2 $.
At high $p_T$ this quadratic growth overwhelms the 
(neglected) linear response due to   $u_{T}^{(4)}$, and leads to the characteristic relation
$v_4 = \frac{1}{2} v_2^2$. An entirely identical argument 
shows that $v_5 = v_2 v_3$  at high momentum in  ideal 
hydrodynamics.

The Borghini-Ollitrault  argument given above shows that the response coefficients  
in ideal hydro should asymptote at large momentum,
\st
  \frac{w_{4(22)}/\epsilon_2^2}
{ (w_{2}/\epsilon_2)^2 }
  \xrightarrow[{p_T\rightarrow \infty} ]{} \frac{1}{2} \,  ,
  \qquad 
  \frac{w_{5(23)}/(\epsilon_2 \epsilon_3) }
  { (w_{2}/\epsilon_2) (w_{3}/\epsilon_3) }
  \xrightarrow[{p_T\rightarrow \infty} ]{} 1 \,  .
\stp
When fluctuations are included these asymptotic relations are modified \cite{Gombeaud:2009ye}:
\st
\label{fluctsv4v5}
  \frac{v_4\{2\}}{v_2\{2\}^2 } \xrightarrow[{p_T\rightarrow \infty} ] { }  
\frac{1}{2} \left(  \frac{ \llangle \epsilon_2^4 \rrangle }{\llangle \epsilon_2^2 \rrangle^2 }
 \right)^{1/2} \, , 
 \qquad
  \frac{v_5\{2\}}{v_2\{2\} v_3\{2\} } \xrightarrow[{p_T\rightarrow \infty} ] { }  
\left(  \frac{ \llangle (\epsilon_2\epsilon_3)^2 \rrangle }{\llangle \epsilon_2^2 \rrangle  \llangle \epsilon_3^2 \rrangle}
 \right)^{1/2} \, . 
\stp
Previous studies of $v_4$ in the $\Psi_2$ plane (see \Sect{vis_dep}) 
have shown that such geometrical factors are essential  
to reproducing the centrality dependence of $v_4/v_2^2$ \cite{Gombeaud:2009ye}. 
The following table records the geometrical ratios in \Eq{fluctsv4v5} 
as a function of centrality. 
\begin{center}
\begin{tabular}{|l|c|c|c|c|c|c|c|c|r}
\hline
Centrality \% & 2.5 & 7.5 & 12.5 & 17.5 & 25.0  & 35.0 & 45.0 & 55.5 \\
\hline
$\sqrt{ { \llangle \epsilon_2^4 \rrangle }/{\llangle
 \epsilon_2^2 \rrangle^2 }}$
& 1.40 & 1.33 & 1.26 & 1.22 & 1.20 & 1.18 & 1.17 & 1.16  \\
\hline
$\sqrt{{ \llangle (\epsilon_2\epsilon_3)^2 \rrangle }/({\llangle
\epsilon_2^2 \rrangle  \llangle \epsilon_3^2 \rrangle})}$
& 0.99 & 0.97 & 0.96 & 0.95 & 0.94 & 0.93 & 0.930 & 0.92 \\
\hline
$\sqrt{{ \llangle \epsilon_2^2 \rrangle }/{\llangle
\epsilon_1^2 \rrangle}}$
& 2.12 & 2.78 & 3.22 & 3.46 & 3.56 & 3.40 & 3.04 & 2.64  \\
\hline
\end{tabular}
\end{center}

\begin{figure}
\includegraphics[width=0.49\textwidth]{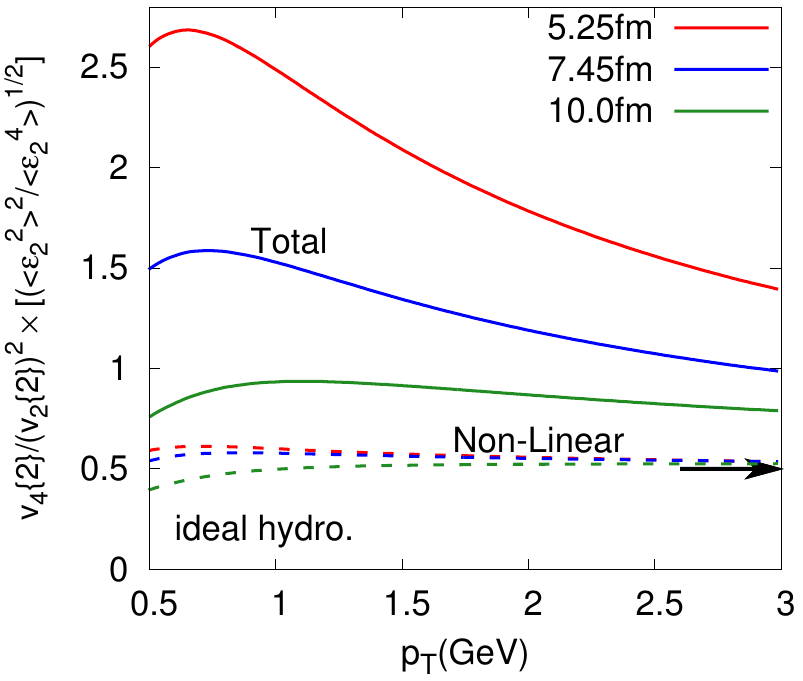}
\includegraphics[width=0.49\textwidth]{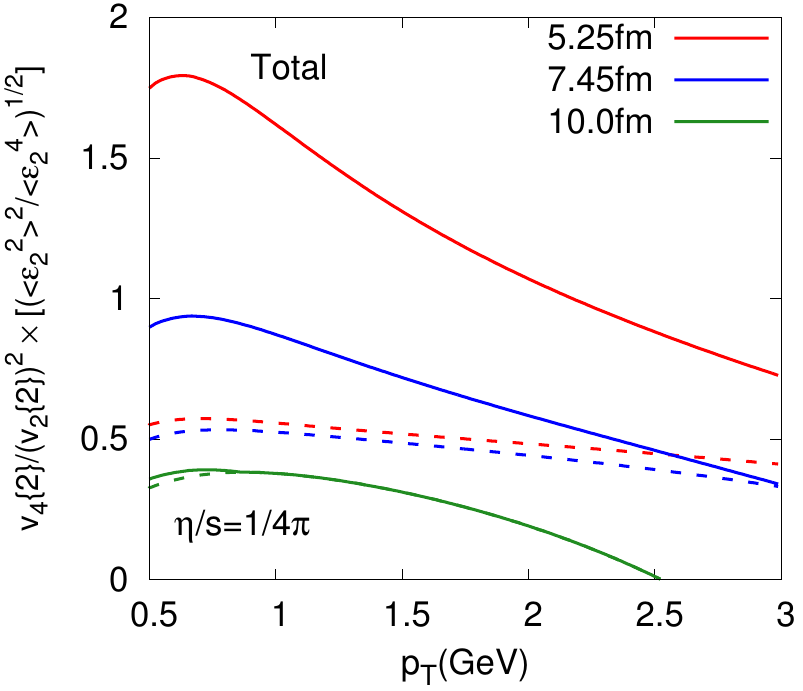}
\includegraphics[width=0.49\textwidth]{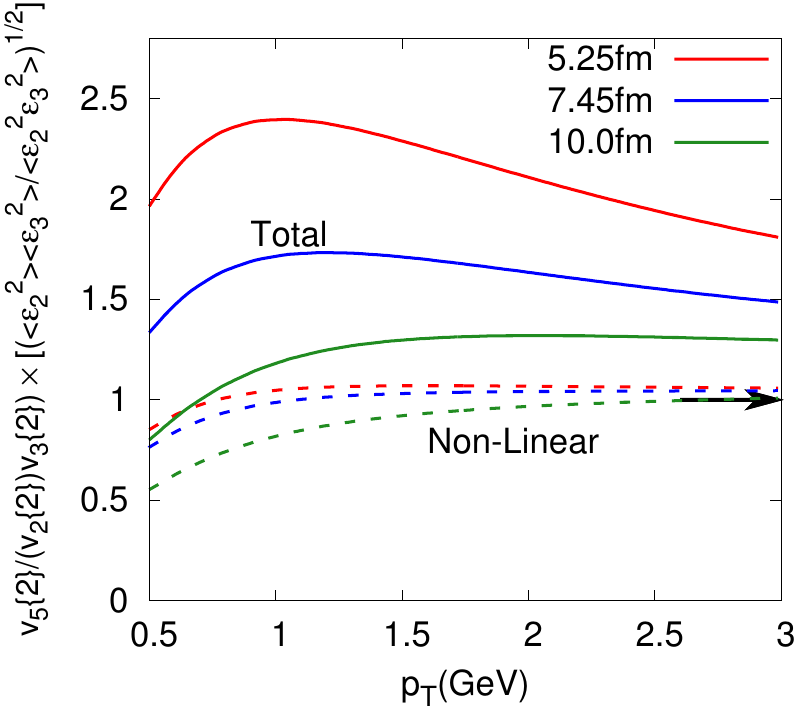}
\includegraphics[width=0.49\textwidth]{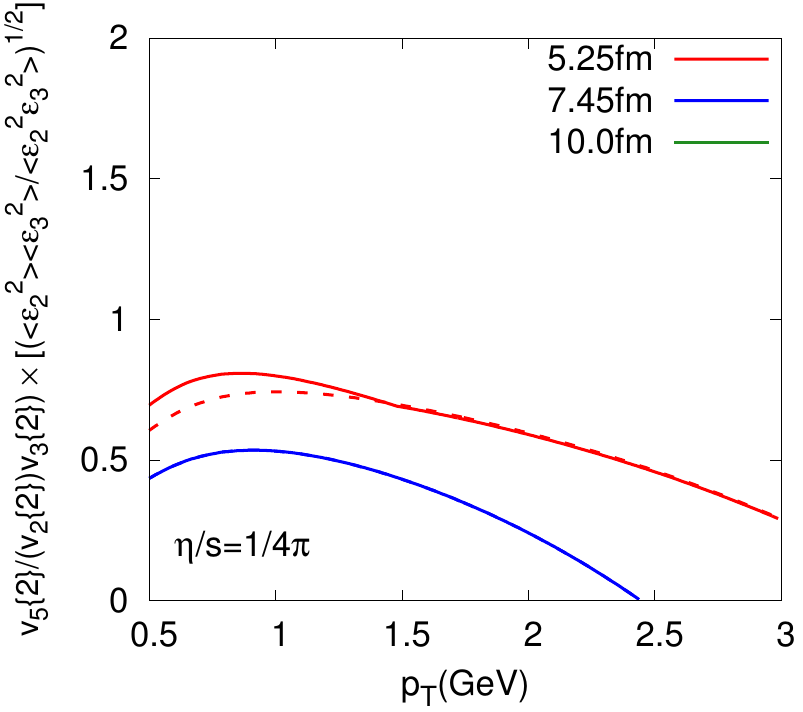}
\caption{ 
Results for $v_4$ and $v_5$ for ideal and viscous hydrodynamics
at various impact parameters. 
The Borghini-Ollitrault expectation is indicated by the arrows for the 
ideal $v_4$ and $v_5$ curves \cite{Borghini:2005kd}.
\label{v4v5scaling_pt}
}
\end{figure}

We have found that rather large $p_T$ is needed to see the  non-linear limit
given by \Eq{fluctsv4v5}. In the current framework,  the linear and non-linear
response terms, and their interference, determine the full result
\st
v_{4}\{2\}(p_T) = \frac{\Dlangle w_4(p_T)w_4 + w_{4(22)}(p_T)w_{4(22)} + 
[w_4(p_T)w_{4(22)} + w_{4(22)}(p_T)w_4]\cos4(\Phi_4-\Phi_2)\Drangle}{v_4\{2\}}
\stp
\Fig{v4v5scaling_pt} shows the complete result for $v_4\{2\}/v_2\{2\}^2$ 
(scaled by $ \sqrt{\llangle \epsilon_2^4 \rrangle/\llangle \epsilon_2^2 \rrangle^2}$)
for ideal and viscous hydrodynamics. Focusing on the 
ideal results, we see that full results (the solid lines)
approach the non-linear expectation  of Borghini and Ollitrault (the dashed
line)  only very slowly. This is in large part because $w_{4}(p_T)$ is only
qualitatively linear at sub-asymptotic $p_T$ and increases almost quadratically
at intermediate $p_T \sim 1.5\,{\rm GeV}$, momentarily keeping up with 
the non-linear response. When viscous corrections are included, the non-linear
results become dominant in peripheral collisions. 
Similar results for $v_5$ 
in ideal and viscous hydrodynamics are also shown in \Fig{v4v5scaling_pt}.
In the viscous case, the non-linear result gives almost the full $v_5\{2\}$ 
for all centrality classes shown.

It is worth noting that the magnitude of the viscous corrections as a function of $p_T$
for $v_4$ and $v_5$ are sensitive to ansatz used for the viscous distribution
function, $\delta f$  \cite{Luzum:2010ad}. In particular, the quadratic ansatz  
used in this work  assumes that the quasi-particle energy loss is
independent of momentum, $dp/dt \propto \mbox{const}$.
A linear ansatz for $\delta f$ is better motivated for QCD like theories and
results in smaller viscous corrections for $v_4$ and $v_5$ as a function of $p_T$ \cite{Dusling:2009df}.
A complete discussion of this point is reserved for future work.

\begin{figure}
\includegraphics[width=0.49\textwidth]{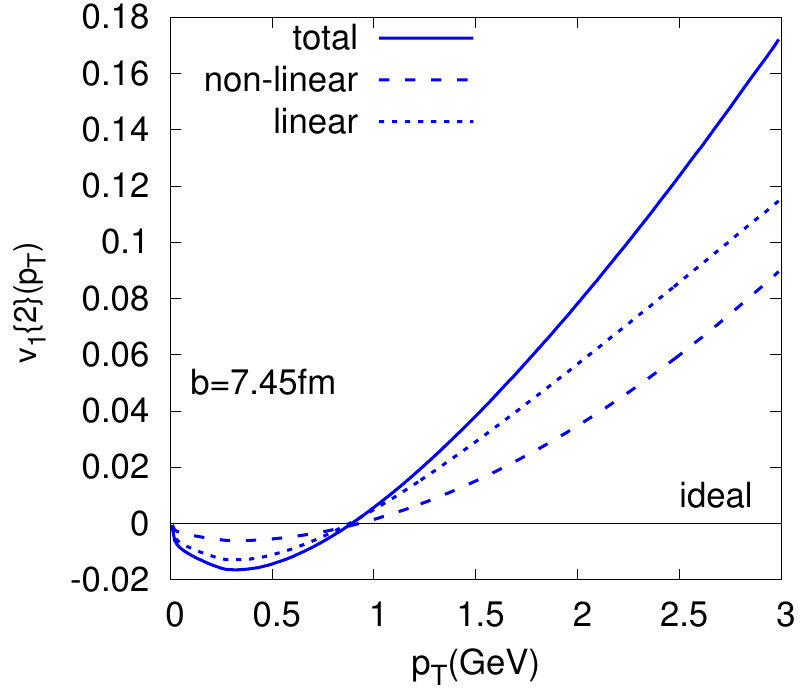}
\includegraphics[width=0.49\textwidth]{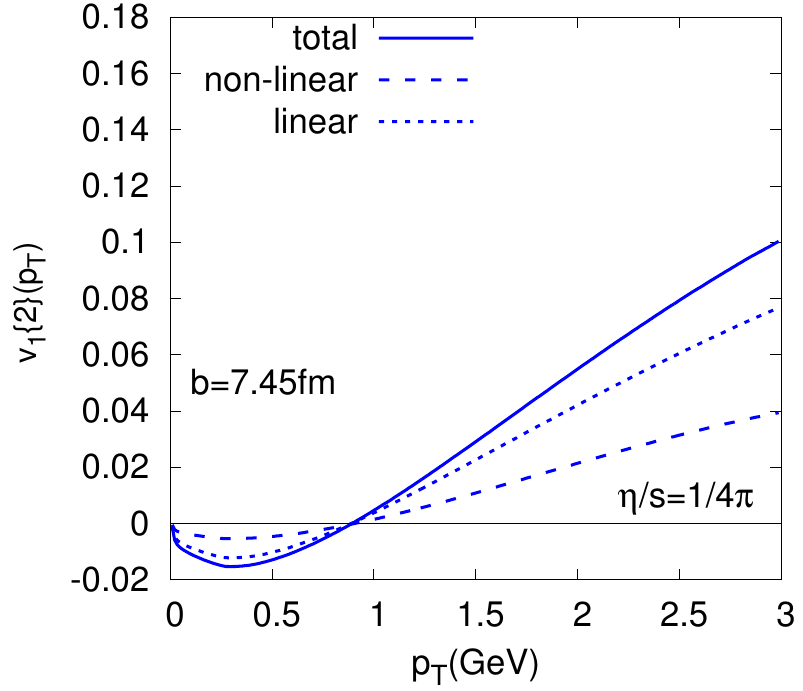}
\caption{ 
   $v_{1}(p_T) \{2\}$ (\Eq{vnptdef}) in ideal and viscous hydrodynamics from
   the linear response to $\e_1$, the  non-linear response to $\e_2\e_3$, and
   the total response, \Eq{numeratorv1}. 
\label{v1scaling_pt}
}
\end{figure}

\Fig{v1scaling_pt} presents the corresponding analysis for $v_1(p_T)$.
We see that the non-linear terms provide a  correction 
to the linear response which grows with $p_T$ due to the quadratic 
dependence of the non-linear  response coefficients, $w_{1(23) } \propto p_T^2$.
We note that the viscous corrections 
are approximately  the same for $v_1(p_T)$ and $v_2(p_T)$, as 
expected from the discussion of viscous corrections
given in \Sect{vis_dep}.

\subsection{Discussion}

We have presented a framework of non-linear response to understand 
the higher harmonics generated in heavy ion collisions.  Then we extracted
the non-linear response coefficients using ideal and viscous hydrodynamics
and studied the dependence on the shear viscosity, in \Fig{response_fig}.
The pattern of viscous corrections is further analyzed in \Fig{visc_response} 
and explained in \Sect{Hydroresponse}.  Generally, when 
the harmonic order is large, the non-linear response
is less damped than the corresponding linear response.  Thus, 
when viscosity is included in hydrodynamic simulations, the non-linear response
becomes increasingly important for the higher harmonics.
This qualitative reasoning is confirmed in \Fig{vn_fig} 
which shows $v_1$, $v_4$ and $v_5$ using linear and non-linear response 
and is the principal result of this work.
We see that the non-linear response is essential for $v_4$ and $v_5$,  and
constitutes an important correction for $v_1$. 

Experimentally,
the relative contributions of the linear and non-linear response can
be disentangled by measuring  $v_5$ in the $2\Psi_2+3\Psi_3$ 
and $\Psi_5$ planes, $i.e.$ by measuring
\st
v_{5(23)} \equiv \llangle \cos (5\phi_\p - 2 \Psi_2 - 3\Psi_3) \rrangle \, ,
   \quad \mbox{and} \quad v_{5(5)} \equiv \llangle \cos5(\phi_\p - \Psi_5) \rrangle \, .
\stp
Although a full discussion of this and similar measurements is reserved for 
future work, a qualitative expectation based on \Fig{vn_fig}(e) and (f) is that
the $\llangle \cos(5\Psi_5 - 3\Psi_3 - 2\Psi_2) \rrangle$ correlation should
be strong compared to the geometric average, and should  change rapidly
from central to mid-central collisions. Qualitatively, this is precisely 
what was observed recently by the ATLAS collaboration \cite{ATLASCorrelations}.

The non-linear response can also be studied by analyzing the $p_T$ dependence
of the flow harmonics.  \Fig{v4v5scaling_pt} and \Fig{v1scaling_pt} exhibit
$v_4(p_T)$,  $v_5(p_T)$, and $v_1(p_T)$.  In ideal
hydrodynamics at large $p_T$ we expect to find $v_4 = \half v_2^2 $ on an event
by event basis. Our non-linear response coefficients corroborate this non-linear
expectation for $v_4$ and an analogous relation for $v_5$, $v_5=v_2 v_3$.  
However, since what is normally measured is $v_4\{2\}/(v_{2}\{2\})^2$
and not $\Dlangle v_4/v_2^2 \Drangle$, this ideal non-linear expectation
must be multiplied by 
$(\llangle \epsilon_2^4 \rrangle / \llangle\epsilon_2^2 \rrangle^2)^{1/2}$ when comparing to the experimental data \cite{Gombeaud:2009ye}. In addition, 
this expectation of ideal hydrodynamics is 
broken by viscous corrections, and  by the linear response to 
the fourth order cumulant  $\c_4$ ({\it i.e.} $\epsilon_4$). 
When all of these corrections are taken 
into account, we find that relations such as $v_4 =\half v_2^2$ and $v_5 = v_2 v_3$ provide only a rough guide to  the full result. 

Throughout we have assumed perfect correlation
between $\Psi_2$ and $\psie_2$ and $\Psi_3$ and $\psie_3$. This strict 
correlation is only approximately true. For instance the combination 
of a $v_1$ and a $v_3$ can yield a $v_2$, 
\st
v_2 e^{-i2\Psi_2}  = w_{2} e^{-i2\psie_2} + w_{2(13)} e^{-i3\psie_3 + i\psie_1 }  \, .
\stp
This naturally provides a correlation between the $\Psi_2$ and $\Psi_3$ plane,
although the geometric correlation between $\psie_2$ and $\psie_3$ is
negligibly small. Indeed the $(\Psi_2,\Psi_3)$ correlation, which was very
recently observed by the ATLAS collaboration \cite{ATLASCorrelations},  is too large to be easily
explained with the geometric correlations of the Glauber model.
Similarly, assuming that the linear response to $\epsilon_6$ is negligible, one
could expect that in central collisions $v_6$ is determined by the quadratic response to
$v_3$, while in peripheral collisions $v_6$ is determined by a cubic response to $v_2$ 
\st
v_6 e^{-i6\Psi_6} = w_{6(222)}e^{-i6\psie_2} +   w_{6(33)} e^{-i6\psie_3}  \, .
\stp
Qualitatively, this pattern is consistent with the observed $(\Psi_6,\Psi_3)$ and
$(\Psi_6,\Psi_2)$ correlations presented in \cite{ATLASCorrelations}.  It will be
interesting to see if all of the observed correlations can be quantitatively
understood with the non-linear response theory outlined in this paper. A full
quantitative comparison with the experimental data is
reserved for future work.


\vspace{\baselineskip}
\noindent{ \bf Acknowledgments:} \\
{}\\
We thank J.~Y. Ollitrault, and Z.~Qiu, and U.~Heinz for many constructive and insightful comments.
D.~Teaney is a RIKEN-RBRC fellow.  This work is supported in part by the
Sloan Foundation and by the Department of Energy through the Outstanding Junior
Investigator program, DE-FG-02-08ER4154.

\bibliography{nlin}

\end{document}